\documentstyle[12pt]{article}
\bibliography{plain}
\title{\bf The relations between d-dimensional isotropic oscillator and D-dimensional like-hydrogen atom}
\author{ Z. Neshati, Z. Bakhshi\thanks{Corresponding author
(E-mail: z.bakhshi@shahed.ac.ir)}
   \\
{\small Department of Physics, Faculty of Basic Sciences, Shahed
University, Tehran, Iran.} \\  }\pagebreak
\begin{document}
\maketitle
\begin{abstract}
Being comparable in quantum systems makes it possible for spaces with varying dimensions
 to attribute each other using special conversions can attribute schr\"{o}dinger
  equation with like-hydrogen atom potential in defined dimensions to a schr\"{o}dinger equation
  with other certified dimensions with isotropic oscillator potential.
   Applying special transformation provides a relationship between
   different dimensions of two quantum systems. The result of the
  quantized isotropic oscillator can be generalized to like-hydrogen atom problem in different dimensions.
  The connection between coordinate spaces in different dimensions can follow a specific relation
  that by using it and applying the parametric definition in two problems, energy spectrum
   and like-hydrogen atom potential wave functions problem will be solved.
\end{abstract}
\pagebreak \vspace{7cm}
\pagebreak \vspace{7cm}
\section{Introduction}

The hidden dynamical symmetry in some quantum systems and the beauty of solving such
issues causes imputation of other quantum systems solvable
 models by using some special transformations.
The solvability of such systems provides enough motivations
 for simulation of other physical systems with them.
Such solvable quantum systems are included in quantum isotropic
oscillators with dimensions 2, 4, 8, 16, 32 that can be attributed
 to like-hydrogen atom quantum systems with dimensions 2, 3, 5, 9.
 The type of transformation is so that special dimensions of two
 quantum systems can be related to each other. Therefore, these
  transformations can't be useful for any dimension of considered
  quantum system.
These transformations create a duality between discrete quantum
 systems named oscillators with different dimensions and continues
  quantum systems that is called like-hydrogen atom quantum systems.
These transformations help attributing solvability of a quantum
 system to the solvability of desired quantum system,
 so different algebra methods
that are applied in solvable models can be expanded in the desired problems.
This expansion causes parametrical relations between two quantum systems.\\
Algebra method such as dynamic symmetry groups corresponded with the
 like-hydrogen atom can be related to the dynamic symmetry groups of
  isotropic oscillators [1]. Also, the explicit form of the transformations binding like-hydrogen
atom wave function and isotropic oscillator wave function can be
available by using the definitions of parameters in two considered systems [2-4].\\
Since it is known that the quantized oscillator is a principle
concept model in the quantum field theory and the problem of D-dimensional oscillator can be related to the
  problem of d-dimensional like-hydrogen atom, by using special transformations,
   so many well developed methods applied in quantum field theory and
   nuclear physics can be also effectively used in the investigations
   of the behavior of like-hydrogen atom in the
  external electric and magnetic fields [5-8].
There are some special transformations that can convert
d-dimensional like-hydrogen atom to D-dimensional isotropic oscillator.
These bilinear transformations are called Levi-Civita [9],
 kustaan- Steif [10],Hurwitz transformations [11] that can depend the dimensions
 of d=2, 3, 5, 9, 17 to the dimensions of D=2, 4, 8, 16, 32, correspondingly.
Since, the above mentioned transformations connected
two different coordinate spaces, dimensionally, so there isn't an one by one relation
between two spaces. The transformations that is applied to changing two spaces with
different dimensions are called non-bijective quadratic transformations [12].
It should stressed that non-bijectivity of the $R^{d}\rightarrow R^{D}$
transformation means that for each element of $R^{d}$,
 there is a whole set of elements in $ R^{D}$ .\\
In this paper, by special transformations, a general
 model is presented that shows how physical quantities
 can be converted to each other in two different
 quantum systems with two different dimensions.
In this method, D-dimensional isotropic
oscillator energy can be  expanded  to d-dimensional
like-hydrogen atom energy by considering the parametrical relations between two systems.
Also, it is shown that wave function of
isotropic oscillator can be written based on wave function of like- hydrogen atom.
The presented general method is applied in 5 and
16 dimensional isotropic oscillators in the sections (3) and (4), respectively.

\section{ Generalized transformations in two different spaces of quantum systems}
\setcounter{equation}{0}

As mentioned  before, transformations of dimensions between two
systems as $d\rightarrow D$ are included only in some cases of dimensions
such as d=2, 3, 5, 9, 17 to the dimensions of D=2, 4, 8, 16, 32, correspondingly.
In general form, there is a like-hydrogen atom with a certain dimension
that can be related to isotropic oscillator with another specific dimension.
Dimensional relations between two spaces $R^{d}$ and $R^{D}$ is
considered by special transformation that only convert $d=2^{m}$
dimension of the isotropic oscillator to  $D=2^{m-1}+1$
dimension of like-hydrogen, where $m$ is an integer
number and $ m=1, 2, 3, 4, 5$ based on restrict of $d$ and $D$.\\
The most important feature of non-bijective quadratic
transformations is the validity of the Euler identity [13]:

\begin{equation}
u_{1}^{2}+u_{2}^{2}+...+u_{d}^{2}=x_{1}^{2}+x_{2}^{2}+...+x_{D}^{2},
\end{equation}

Where $x_{i}$ and $u_{k}$ are the D-dimensional like-hydrogen
atom and d-dimensional isotropic oscillator coordinates, respectively.\\
Introducing non-bijective quadratic transformation, coordinates $(x_{1}, ..., x_{D})$ of space $R^{D}$ can be related to coordinates
$(u_{1}, ..., u_{d})$ of the space $R^{d}$, by [12]:
\begin{equation}
x_{i}=\sum_{j=1}^{d}T_{ij}u_{j},\qquad\qquad\qquad i=1, 2, 3, ..., D
\end{equation}
where $x_{i}$ is a generalized coordinate in space $R^{D}$ and $T_{ij}$ is a transformation
matrix that shows the type of transformations applied to the problem.
\\Transformation matrix has basic properties. First, the elements are linear homogeneous functions of $u_{j}$.
 Second, matrix is orthogonal in the following sense:\\
(a) The scalar produces two different rows vanishes.\\
(b) Each row has the norm $u_{1}^{2}+u_{2}^{2}+...+u_{d}^{2}$.\\
Multiplying two transformation matrixes and summing over $j$ based on the Euler identify (2.1), it can be shown that [12]:
\begin{equation}
\sum_{j=1}^{d}T_{ij}T_{kj}=u^{2}\delta_{ik}, \qquad\qquad\qquad i, k=1, 2, ..., D,
\end{equation}
and
\begin{equation}
\sum_{j=1}^{d}\frac{\partial T_{ij}}{\partial u_{j}}=0.
\end{equation}
The transformation matrixes $T_{ij}$ should be defined so that:
\begin{equation}
\frac{\partial x_{i}}{\partial u_{j}}=2T_{ij},\qquad\qquad\qquad i=1,...,D.
\end{equation}
Considering an arbitrary function of $u$ as $\tau(u)$ and using equation (2.5):
\begin{equation}
\frac{\partial\tau(u) }{\partial
u_{j}}=2\sum_{k=1}^{D}T_{kj}\frac{\partial\tau(u) }{\partial x_{k}},\qquad\qquad\qquad j=1, 2, ..., d.
\end{equation}
Multiplying (2.6) by $T_{ij}$ and using relation (2.3), it can be easily shown that:
\begin{equation}
\frac{\partial\tau(u)}{\partial
x_{i}}=\frac{1}{2u^{2}}\sum_{j=1}^{d}T_{ij}\frac{\partial\tau(u)}{\partial u_{j}}.
\end{equation}
Therefore, relations (2.6) and (2.7) can be generalized by:
\begin{equation}
\frac{\partial}{\partial
x_{i}}=\frac{1}{2u^{2}}\sum_{j=1}^{d}T_{ij}\frac{\partial}{\partial u_{j}},
\end{equation}
\begin{equation}
\frac{\partial}{\partial
u_{j}}=2\sum_{k=1}^{D}T_{kj}\frac{\partial}{\partial x_{k}},
\end{equation}
If $\varphi_{i}$ operators are introduced as:

\begin{equation}
\hat{\varphi}_{i}=iu^{2}\frac{\partial}{\partial
q_{i}}=\frac{i}{2}\sum_{j=1}^{d}T_{ij}\frac{\partial}{\partial u_{j}},\qquad\qquad\qquad i=D, D+1, ..., d-1, d.
\end{equation}
Considering relation (2.5), $\varphi_{i}$ will operate as follows:
\begin{equation}
\hat{\varphi}_{i}\hat{x}_{k}=iu^{2}\delta_{ik}=0,\qquad\qquad i=D,..., d,\qquad\qquad k=1, ..., D.
\end{equation}
Also, it is clear that relation (2.11) is valid for an arbitrary functions of $x$, so that $\hat{\varphi}_{i}F(x)=0$.
Considering differential relations between two different coordinates $x_{i}$ and $u_{j}$,
the relation between like-hydrogen atom and isotropic oscillator Laplacians is as follows [12]:
\begin{equation}
\Delta_{x}=\frac{1}{4u^{2}}\Delta_{u}+\frac{1}{u^{4}}\varphi^{2},
\end{equation}
where operator $\hat{\varphi}^{2}$ is defined as:
\begin{equation}
\hat{\varphi}^{2}=\hat{\varphi}_{D}^{2}+ ... +\hat{\varphi}_{d}^{2}.
\end{equation}
 $\hat{\varphi}^{2}$ is the operator that causes laplacian of like-hydrogen atom, written based on the laplacian of isotropic oscillator.
 Therefore, D-dimensional Schr\"{o}dinger equation for like-hydrogen atom as:
\begin{equation}
(-\frac{\hbar^{2}}{2\mu}\triangle_{x}-\frac{e^{2}}{r})\psi(x)=E\psi(x),
\end{equation}
can be transformed into the form:
\begin{equation}
(-\frac{\hbar^{2}}{2\mu}\triangle_{u}-4Eu^{2})\psi(x)=4e^{2}\psi(x).
\end{equation}
Since equation (2.15) is comparable with solvable Schr\"{o}dinger equation for d-dimensional isotropic oscillator as:
\begin{equation}
(-\frac{\hbar^{2}}{2\mu}\triangle_{u}+\frac{1}{2}\mu\omega^{2}u^{2})\varphi(u)=\varepsilon\varphi(u).
\end{equation}
Thus, equation (2.15) that is related to D-dimensional like-hydrogen atom, can be
solved considering the solution of d-dimensional isotropic oscillator.
Applying d-dimensional isotropic oscillator as a solvable model will be perfect,
if the following parametrical relations is satisfied between two equations (2.15) and (2.16):
\begin{equation}
\frac{1}{2}\mu\omega^{2}=-4E,\qquad\qquad\qquad 4e^{2}=\varepsilon.
\end{equation}
The energy spectrum in d-dimensional isotropic oscillator can be obtained as:
\begin{equation}
\varepsilon=\hbar\omega(N+2^{m-1}),\qquad\qquad N=0, 1, 2, ... ,\qquad\qquad m=1, 2, 3, ... .
\end{equation}
where $m$ is an integer number and $N$ is a main quantum number that represents the number
of energy discrete bound states levels. This energy discrete spectrum is corresponded with isotropic oscillator in the $2^{m}$ dimensions of space $R^{d}$.
It should be mentioned that in equation (2.16), $\omega$ is considered as
a potential constant parameter, so, $\varepsilon$ is quantized according to relation (2.18).
Although, in equation (2.15), $\varepsilon=4e^{2}$ is assumed as a potential constant parameter
and $\omega$ is quantized by $\frac{\varepsilon}{\hbar(N+2^{m-1})}$ where $N$  is a natural number.
Substituting parametrical relations (2.17) in relation (2.18), energy discrete spectrum of D-dimensional like-hydrogen atom can be calculated by:
\begin{equation}
E=\frac{-2\mu e^{4}}{\hbar^{2}(N+2^{m-1})^{2}},
\end{equation}
where $N$ as a number of energy bound states level can be introduced by
parametrical definitions in the like-hydrogen atom systems with dimensions of $2^{m-1}+1$.\\
According to the type of transformations in the problem $\psi(x)$ is an
even function of variables $u$, so that $\psi(x(u))=\psi(x(-u))$. Therefore, $\psi(x)$ as the
solution of equation (2.14) can be expanded in a full system of even solutions $\varphi_{N_{\alpha}}$ of equation (2.16) [12]:
\begin{equation}
\psi_{n}(x)=\sum C_{n_{\alpha}}\varphi_{N_{\alpha}}(u),
\end{equation}
where $\alpha$ denotes all other quantum numbers.

\section{$ks$ transformations in the four-dimensional quantum isotropic oscillator}
\setcounter{equation}{0}
Presented general model can be used for expansion of four-dimensional
like-hydrogen atom by $Ks$ transformations. The $Ks$ transformations is presented by following matrix [10]:
\begin{equation}
T_{ij}=\left(\begin{array}{cccc}
        u_{1} & -u_{2} & -u_{3} & u_{4}\\
         u_{2} & u_{1} & -u_{4} & u_{3}\\
          u_{3} & -u_{4} & u_{1} & u_{2}\\
           u_{4} & -u_{3} & -u_{2} & -u_{1}\\
              \end{array}\right),
\end{equation}
where mentioned before, basic properties. According to the relation (2.2), connection between $x_{i}$ coordinates of three-dimensional like-hydrogen atom
and $u_{j}$ coordinates of four-dimensional isotropic oscillator is arranged as follows:
\begin{eqnarray}
x_{1}=u_{1}^{2}-u_{2}^{2}-u_{3}^{2}+u_{4}^{2}\nonumber
\\x_{2}=2(u_{1}u_{2}-u_{3}u_{4})\nonumber\\x_{3}=2(u_{1}u_{3}+u_{2}u_{4})
\end{eqnarray}
The $\hat{\varphi}_{i}$ operator has only one form as $\hat{\varphi}_{4}$ that is defined based on relation (2.10) as:
\begin{equation}
\hat{\varphi}_{4}=\frac{i}{2}(u_{4}\frac{\partial}{\partial u_{1}}-u_{3}\frac{\partial}{\partial u_{2}}+u_{2}\frac{\partial}{\partial u_{3}}-u_{1}\frac{\partial}{\partial u_{4}}),
\end{equation}
The $\hat{\varphi}^{2}$ operator can be written based on
$\hat{\varphi}_{4}^{2}$ operator by considering relation (2.14):
\begin{equation}
\hat{\varphi}^{2}=\hat{\varphi}^{2}_{4}.
\end{equation}
Therefore, schr\"{o}dinger equation of like-hydrogen atom in three dimensions
can be comparable with schrodinger equation of isotropic oscillator in four dimensions.
Considering parametrical relations (2.17), $\varepsilon$ energy spectrum
 in four-dimensional isotropic oscillator as:
\begin{equation}
\varepsilon=\hbar\omega(N+2).
\end{equation}
where $m=2$ and $N$ is a level number of energy spectrum. This energy spectrum can be
applied to calculation of energy discrete spectrum of three-dimensional like-hydrogen atom.
 According to relation (2.20), if the integer number $m$ is considered as $m=2$, energy spectrum of like-hydrogen atom in three-dimensions as follows:
\begin{equation}
E=\frac{-2\mu e^{4}}{\hbar^{2}(N+2)^{2}},
\end{equation}
where $N$ parameter can be written by parametrical definitions for three-dimensional like-hydrogen atom.

\section{$Hurwitz$ transformations in the sixteen-dimensional quantum isotropic oscillator}
The transformation of nine-dimensional like-hydrogen atom to sixteen-dimensional isotropic oscillator
will be followed by mentioned general model, if Hurwitz transformation is applied as [11]:
\begin{equation}
\tiny{T_{ij}=\left(\begin{array}{cccccccccccccccc}
        u_{9} & -u_{10} & -u_{11} & u_{12}&-u_{13}&u_{14}&-u_{15}&u_{16}&u_{1}&u_{2}&u_{3}&u_{4}&-u_{5}&u_{6}&-u_{7}&u_{8}\\
          -u_{10} & -u_{9} & -u_{12} & u_{11}&-u_{14}&-u_{13}&-u_{16}&u_{15}&u_{2}&-u_{1}&u_{4}&u_{3}&-u_{6}&-u_{5}&-u_{8}&u_{7}\\
          u_{11} & -u_{12} & u_{9} & u_{10}&u_{15}&-u_{16}&u_{13}&u_{14}&u_{3}&u_{4}&u_{1}&-u_{2}&-u_{7}&u_{8}&u_{5}&-u_{6}\\
           -u_{12} & -u_{11}&-u_{10} & u_{9}&u_{16}&u_{15}&u_{14}&u_{13}&u_{4}&-u_{3}&-u_{2}&-u_{1}&u_{8}&u_{7}&u_{6}&u_{5}\\
           u_{13} & u_{14}&-u_{15} &-u_{16}&u_{9}&-u_{10}&u_{11}&u_{12}&u_{5}&-u_{6}&u_{7}&u_{8}&u_{1}&u_{2}&-u_{3}&-u_{4}\\
             u_{14} & u_{13}&u_{16} &-u_{15}&u_{10}&u_{9}&-u_{12}&u_{11}&u_{6}&u_{5}&u_{8}&-u_{7}&u_{2}&-u_{1}&-u_{4}&u_{3}\\
             u_{5} & -u_{16}&u_{13} &-u_{14}&-u_{11}&u_{12}&u_{9}&u_{10}&u_{7}&u_{8}&-u_{5}&u_{6}&u_{3}&-u_{4}&u_{1}&-u_{2}\\
              -u_{16} & -u_{15}&-u_{14} &-u_{13}&u_{12}&-u_{11}&-u_{10}&u_{9}&u_{8}&-u_{7}&-u_{6}&u_{5}&-u_{4}&-u_{3}&-u_{2}&-u_{1}\\
               u_{1} &u_{2}&u_{3} &u_{4}&u_{5}&u_{6}&u_{7}&u_{8}&-u_{9}&-u_{10}&-u_{11}&-u_{12}&-u_{13}&-u_{14}&-u_{15}&-u_{16}\\
               -u_{9} &u_{10}&-u_{11} &-u_{12}&u_{13}&-u_{14}&u_{15}&-u_{16}&u_{1}&-u_{2}&u_{3}&u_{4}&-u_{5}&u_{6}&-u_{7}&u_{8}\\
                u_{10} &u_{11}&u_{12} &u_{13}&u_{14}&u_{15}&u_{16}&u_{9}&-u_{8}&-u_{1}&-u_{2}&-u_{3}&-u_{4}&-u_{5}&-u_{6}&-u_{7}\\
                 -u_{11} &u_{12}&u_{13} &u_{14}&u_{15}&-u_{16}&-u_{9}&u_{10}&u_{7}&-u_{8}&u_{1}&-u_{2}&-u_{3}&-u_{4}&-u_{5}&-u_{6}\\
                 u_{12} &u_{13}&u_{14}&u_{15}&u_{16}&u_{9}&u_{10}&u_{11}&-u_{6}&-u_{7}&-u_{8}&-u_{1}&-u_{2}&-u_{3}&-u_{4}&-u_{5}\\
                 u_{9} &-u_{10}&u_{11}&-u_{12}&-u_{13}&-u_{14}&-u_{15}&u_{16}&-u_{1}&u_{2}&-u_{3}&u_{4}&u_{5}&u_{6}&u_{7}&-u_{8}\\
                  -u_{10}&-u_{11}&-u_{12}&-u_{13}&-u_{14}&-u_{15}&-u_{16}&-u_{9}&u_{8}&u_{1}&u_{2}&u_{3}&u_{4}&u_{5}&u_{6}&u_{7}\\
                  u_{11}&u_{12}&-u_{13}&-u_{14}&u_{15}&u_{16}&-u_{9}&-u_{10}&u_{7}&u_{8}&-u_{1}&-u_{2}&u_{3}&u_{4}&-u_{5}&-u_{6}\\
              \end{array}\right),}
\end{equation}
Hurwitz transformation matrix is arranged so that the basic properties of transformations are satisfied in it.
 $x_{i}$ coordinates of nine-dimensional like-hydrogen atom are written based on the $u_{j}$
 Coordinates of sixteen-dimensional isotropic oscillator by considering relation (2.2):
\begin{eqnarray}
x_{1}=2(u_{1}u_{9}+u_{2}u_{10}-u_{3}u_{11}+u_{4}u_{12}-u_{5}u_{13}+u_{6}u_{14}-u_{7}u_{15}+u_{8}u_{16}),\nonumber
\\x_{2}=2(-u_{1}u_{10}+u_{2}u_{9}+u_{3}u_{12}+u_{4}u_{11}-u_{5}u_{14}-u_{6}u_{13}+u_{7}u_{16}+u_{8}u_{15}),\nonumber
\\x_{3}=2(u_{1}u_{11}-u_{2}u_{12}+u_{3}u_{10}+u_{4}u_{10}+u_{5}u_{15}-u_{6}u_{16}-u_{7}u_{13}+u_{8}u_{14}),\nonumber
\\x_{4}=2(-u_{1}u_{12}-u_{2}u_{11}-u_{3}u_{10}+u_{4}u_{9}+u_{5}u_{16}+u_{6}u_{15}+u_{7}u_{14}+u_{8}u_{13}),\nonumber
\\x_{5}=2(u_{1}u_{13}+u_{2}u_{14}-u_{3}u_{15}-u_{4}u_{16}+u_{5}u_{9}-u_{6}u_{10}+u_{7}u_{11}+u_{8}u_{12}),\nonumber
\\x_{6}=2(-u_{1}u_{14}+u_{2}u_{13}+u_{3}u_{16}-u_{4}u_{15}+u_{5}u_{10}+u_{6}u_{9}-u_{7}u_{12}+u_{8}u_{11}),\nonumber
\\x_{7}=2(u_{1}u_{15}-u_{2}u_{16}+u_{3}u_{13}-u_{4}u_{14}-u_{5}u_{11}+u_{6}u_{12}+u_{7}u_{9}+u_{8}u_{10}),\nonumber
\\x_{8}=2(-u_{1}u_{16}-u_{2}u_{15}-u_{3}u_{14}-u_{4}u_{13}-u_{5}u_{12}-u_{6}u_{1}-u_{7}u_{10}+u_{8}u_{9}),\nonumber
\\x_{9}=u_{1}^{2}+u_{2}^{2}+u_{3}^{2}+u_{4}^{2}+u_{5}^{2}+u_{6}^{2}+u_{7}^{2}+u_{8}^{2}-u_{9}^{2}-u_{10}^{2}-u_{11}^{2}-u_{12}^{2}-\nonumber
\\u_{13}^{2}-u_{14}^{2}-u_{15}^{2}-u_{16}^{2}.
\end{eqnarray}
$\hat{\varphi}_{i}$ differential operators are written by using relations (2.10) and (4.1):
\begin{eqnarray}
\hat{\varphi}_{10}=\frac{i}{2}(-u_{9}\frac{\partial}{\partial u_{1}}+u_{10}\frac{\partial}{\partial u_{2}}-u_{11}\frac{\partial}{\partial u_{3}}
-u_{12}\frac{\partial}{\partial u_{4}}+u_{13}\frac{\partial}{\partial u_{5}}-u_{14}\frac{\partial}{\partial u_{6}}+u_{15}\frac{\partial}{\partial u_{7}}-u_{16}\frac{\partial}{\partial u_{8}}\nonumber
\\+u_{1}\frac{\partial}{\partial u_{9}}-u_{2}\frac{\partial}{\partial u_{10}}+u_{3}\frac{\partial}{\partial u_{11}}
+u_{4}\frac{\partial}{\partial u_{12}}-u_{5}\frac{\partial}{\partial u_{13}}+u_{6}\frac{\partial}{\partial u_{14}}-u_{7}\frac{\partial}{\partial u_{15}}+u_{8}\frac{\partial}{\partial u_{16}}),\nonumber
\\\hat{\varphi}_{11}=\frac{i}{2}(-u_{10}\frac{\partial}{\partial u_{1}}+u_{11}\frac{\partial}{\partial u_{2}}-u_{12}\frac{\partial}{\partial u_{3}}
-u_{13}\frac{\partial}{\partial u_{4}}+u_{14}\frac{\partial}{\partial u_{5}}-u_{15}\frac{\partial}{\partial u_{6}}+u_{16}\frac{\partial}{\partial u_{7}}-u_{9}\frac{\partial}{\partial u_{8}}\nonumber
\\-u_{8}\frac{\partial}{\partial u_{9}}-u_{1}\frac{\partial}{\partial u_{10}}-u_{2}\frac{\partial}{\partial u_{11}}
-u_{3}\frac{\partial}{\partial u_{12}}-u_{4}\frac{\partial}{\partial u_{13}}-u_{5}\frac{\partial}{\partial u_{14}}-u_{6}\frac{\partial}{\partial u_{15}}-u_{7}\frac{\partial}{\partial u_{16}}),\nonumber
\\\hat{\varphi}_{12}=\frac{i}{2}(-u_{11}\frac{\partial}{\partial u_{1}}+u_{12}\frac{\partial}{\partial u_{2}}+u_{13}\frac{\partial}{\partial u_{3}}
+u_{14}\frac{\partial}{\partial u_{4}}+u_{15}\frac{\partial}{\partial u_{5}}-u_{16}\frac{\partial}{\partial u_{6}}-u_{9}\frac{\partial}{\partial u_{7}}+u_{10}\frac{\partial}{\partial u_{8}}\nonumber
\\+u_{7}\frac{\partial}{\partial u_{9}}-u_{8}\frac{\partial}{\partial u_{10}}+u_{1}\frac{\partial}{\partial u_{11}}
-u_{2}\frac{\partial}{\partial u_{12}}-u_{3}\frac{\partial}{\partial u_{13}}-u_{4}\frac{\partial}{\partial u_{14}}-u_{5}\frac{\partial}{\partial u_{15}}+u_{6}\frac{\partial}{\partial u_{16}}),\nonumber
\\\hat{\varphi}_{13}=\frac{i}{2}(u_{12}\frac{\partial}{\partial u_{1}}+u_{13}\frac{\partial}{\partial u_{2}}+u_{14}\frac{\partial}{\partial u_{3}}
+u_{15}\frac{\partial}{\partial u_{4}}+u_{16}\frac{\partial}{\partial u_{5}}+u_{9}\frac{\partial}{\partial u_{6}}+u_{10}\frac{\partial}{\partial u_{7}}+u_{11}\frac{\partial}{\partial u_{8}}\nonumber
\\-u_{6}\frac{\partial}{\partial u_{9}}-u_{7}\frac{\partial}{\partial u_{10}}-u_{8}\frac{\partial}{\partial u_{11}}
-u_{1}\frac{\partial}{\partial u_{12}}-u_{2}\frac{\partial}{\partial u_{13}}-u_{3}\frac{\partial}{\partial u_{14}}-u_{4}\frac{\partial}{\partial u_{15}}-u_{5}\frac{\partial}{\partial u_{16}}),\nonumber
\\\hat{\varphi}_{14}=\frac{i}{2}(u_{9}\frac{\partial}{\partial u_{1}}-u_{10}\frac{\partial}{\partial u_{2}}+u_{11}\frac{\partial}{\partial u_{3}}
-u_{12}\frac{\partial}{\partial u_{4}}-u_{13}\frac{\partial}{\partial u_{5}}-u_{14}\frac{\partial}{\partial u_{6}}-u_{15}\frac{\partial}{\partial u_{7}}+u_{16}\frac{\partial}{\partial u_{8}}\nonumber
\\-u_{1}\frac{\partial}{\partial u_{9}}+u_{2}\frac{\partial}{\partial u_{10}}-u_{3}\frac{\partial}{\partial u_{11}}
+u_{4}\frac{\partial}{\partial u_{12}}+u_{5}\frac{\partial}{\partial u_{13}}+u_{6}\frac{\partial}{\partial u_{14}}+u_{7}\frac{\partial}{\partial u_{15}}-u_{8}\frac{\partial}{\partial u_{16}}),\nonumber
\\\hat{\varphi}_{15}=\frac{i}{2}(-u_{10}\frac{\partial}{\partial u_{1}}-u_{11}\frac{\partial}{\partial u_{2}}-u_{12}\frac{\partial}{\partial u_{3}}
-u_{13}\frac{\partial}{\partial u_{4}}-u_{14}\frac{\partial}{\partial u_{5}}-u_{15}\frac{\partial}{\partial u_{6}}-u_{16}\frac{\partial}{\partial u_{7}}-u_{9}\frac{\partial}{\partial u_{8}}\nonumber
\\+u_{8}\frac{\partial}{\partial u_{9}}+u_{1}\frac{\partial}{\partial u_{10}}+u_{2}\frac{\partial}{\partial u_{11}}
+u_{3}\frac{\partial}{\partial u_{12}}+u_{4}\frac{\partial}{\partial u_{13}}+u_{5}\frac{\partial}{\partial u_{14}}+u_{6}\frac{\partial}{\partial u_{15}}+u_{7}\frac{\partial}{\partial u_{16}}),\nonumber
\\\hat{\varphi}_{16}=\frac{i}{2}(u_{11}\frac{\partial}{\partial u_{1}}+u_{12}\frac{\partial}{\partial u_{2}}-u_{13}\frac{\partial}{\partial u_{3}}
-u_{14}\frac{\partial}{\partial u_{4}}+u_{15}\frac{\partial}{\partial u_{5}}+u_{16}\frac{\partial}{\partial u_{6}}-u_{9}\frac{\partial}{\partial u_{7}}-u_{10}\frac{\partial}{\partial u_{8}}\nonumber
\\+u_{7}\frac{\partial}{\partial u_{9}}+u_{8}\frac{\partial}{\partial u_{10}}-u_{1}\frac{\partial}{\partial u_{11}}
-u_{2}\frac{\partial}{\partial u_{12}}+u_{3}\frac{\partial}{\partial u_{13}}+u_{4}\frac{\partial}{\partial u_{14}}-u_{5}\frac{\partial}{\partial u_{15}}-u_{6}\frac{\partial}{\partial u_{16}}).
\end{eqnarray}
According to relation (2.13), the $\hat{\varphi}^{2}$ operator is defined as follows:
\begin{equation}
\hat{\varphi}^{2}=\hat{\varphi}^{2}_{10}+\hat{\varphi}^{2}_{11}+ ... +\hat{\varphi}^{2}_{16}.
\end{equation}
$\hat{\varphi}^{2}$ is the operator that connects Laplacian of nine-dimensional
like-hydrogen atom to sixteen-dimensional isotropic oscillator.
 Therefore, $\varepsilon$ energy spectrum of sixteen-dimensional isotropic oscillator as:
\begin{equation}
\varepsilon=\hbar\omega(N+8),
\end{equation}
where is obtained from relations (2.18), is connected to energy discrete
spectrum of nine-dimensional like-hydrogen atom. $E$ energy spectrum of
like-hydrogen atom in nine-dimensions, for $m=4$, can be obtained with relation(2.19):
\begin{equation}
E=\frac{-2\mu e^{4}}{\hbar^{2}(N+8)^{2}},
\end{equation}
where $N$ parameter is defined based on like-hydrogen atom in nine-dimensions. Since there exists
a hidden non-abelian monopole in the sixteen- dimensional isotropic oscillator, so, this system looks like
 a nine-dimensional like- hydrogen atom in a field of monopole desired by septet of potential vectors [8].

\section{Conclusion} \setcounter{equation}{0}
Using appropriate transformations and mathematical computations creates
 a relationship between d-dimensional isotropic oscillator and
  D-dimensional like-hydrogen atom coordinations. It is considered that mathematical calculations leads
relationship between D-dimensional laplacian of like-hydrogen atom
and d-dimensional laplacian of isotropic oscillator. Energy spectrum and wave function are calculated through replacing obtained
relations in Schr\"{o}dinger equation of D-dimensional like-hydrogen
atom and comparing of this equation to d-dimensional isotropic
oscillator problems.

\pagebreak \vspace{7cm}
\pagebreak 
\end{document}